\documentclass{icrc2009}
\usepackage{graphicx}   
\usepackage{caption}    
\usepackage[font=footnotesize]{subfig} 
\usepackage{fixltx2e}
\usepackage{url}

\newcommand{\shorttitle}[1]%
{\markboth{Proceedings of the 31\MakeLowercase{$^{st}$} ICRC, {\L}\'{o}d\'{z} 2009}{#1} }
\newcommand{\etal}{\MakeLowercase{\textit{et al. }}} 

\begin{document}
\title{About chemical composition of the primary cosmic radiation at ultra-high energies}

\author{\IEEEauthorblockN{T.M. Roganova\IEEEauthorrefmark{1},
                          L.G. Dedenko\IEEEauthorrefmark{1}
			  A.V. Glushkov\IEEEauthorrefmark{2},
                          G.F. Fedorova\IEEEauthorrefmark{1},
                          S.P. Knurenko\IEEEauthorrefmark{2},
                          I.T. Makarov\IEEEauthorrefmark{2},\\
                          D.A. Podgrudkov\IEEEauthorrefmark{1},
                          M.I. Pravdin\IEEEauthorrefmark{2} and
                          I.Ye. Sleptzov\IEEEauthorrefmark{2}}
                            \\
\IEEEauthorblockA{\IEEEauthorrefmark{1}D.V. Skobeltsyn Institute of Nuclear Physics, MSU,\\ Leninskie Gory,
119992 Moscow, Russia}
\IEEEauthorblockA{\IEEEauthorrefmark{2}Yu.G. Shafer Institute of Cosmophysical Research and Aeronomy, \\
31 Lenin Ave, 677891 Yakutsk, Russia}}

\shorttitle{T.M. Roganova \etal About chemical composition...}
\maketitle

\begin{abstract}

The Yakutsk array includes the surface scintillation detectors and detectors of the Vavilov-Cherenkov radiation
and underground detectors of muons with energies above 1~GeV.
All these detectors readings are suggested to be used to study chemical  composition of the primary cosmic
radiation at ultra-high energies in terms of some model of hadron interactions. The fluxes of electrons,
positrons, gammas, Cherenkov photons and muons in individual extensive air showers induced by the primary
protons and helium, oxygen and iron nuclei at the level of observation have been estimated with help of
the code CORSICA 6.616. The thinning parameter ${10}^{-7}$ have been used. Calculations have been carried
out in terms of the QGSJET-2 and Gheisha-2002 models. The responses of various detectors are estimated
with the help of the code GEANT4. First, energies $E$ and coordinates $X$ and $Y$ of the core of individual
extensive air showers with observed the zenith and azimuth angles have been estimated using all
surface scintillation detector readings instead of using of the standard procedure with a parameter $s(600)$.
These detector readings have been compared with the detector responses, calculated for all particles which
hit the scintillation detectors in each individual shower with observed the zenith and azimuth angles.
This comparison show that the values of the function ${\chi}^{2}$ per one degree of freedom changes from 1.1 for iron
nuclei to 0.9 for primary protons. As this difference is small  all readings of detectors of the
Vavilov-Cherenkov radiation have been used.  At last, readings of underground detectors of muons with
energies above 1~GeV have been exploited to make definite conclusion about chemical composition. The
primary gammas are not  favourable due to large contribution to a signal in the surface scintillation detectors.
\end{abstract}

\begin{IEEEkeywords}
 chemical composition
\end{IEEEkeywords}

\section{Introduction}

The study of the chemical composition of the primary cosmic radiation at ultra high energies is of important.
Decreasing of the flux of the primary protons at energies above $\sim 6\cdot {10}^{19}$eV has been predicted
by Greisen,
Zatsepin and Kuzmin \cite{1,2} (effect GZK) due to interactions of these primary protons with microwave background
radiation. This suppression of the flux of cosmic radiation at the energy mentioned above would not be seen in
case if heavier primaries such as iron nuclei dominate the composition of this cosmic radiation. One more point
of great interest is the presence of the primary photons at such ultra high energies. Due to the GZK effect or
due to some possible top-down scenarios of origin of cosmic rays such primary photons should give some
contribution to the flux of the primary cosmic radiation. Searching for these primary photons has been
resulted in setting some upper limits on the fraction of these photons at various energies \cite{3} -- \cite{8}.
The only
key to success in almost all attempts to study chemical composition is a dependence of the muon number $N_m$ on
energy $E$ of the primary particle which induced an extensive air shower:

\begin{equation}
N_m=a\cdot E^b,             \label{eq:1}
\end{equation}

\noindent
where $a$ and $b$ are constant values and the exponent $b$ is not exceed 1. It is a common agreement then that the
photon induced showers would have smaller fraction of muons relatively to all secondary particles
due to small cross sections of the photonuclear interactions than the showers
induced by the primary protons. And the contrary, the showers induced by the primary iron nuclei would have larger
fraction of muons. Of course, it is a very model dependent point. Nevertheless, many attempts have been made to study
chemical composition by comparing some distributions of number of muons \cite{9} with the appropriate data
\cite{10} -- \cite{12}. Many
other suggestions have been done to use time distributions of muons, their height production distributions, $X_{max}$
distribution and so on.
At ultra high energies the only variables which can be used to study chemical composition are the height $X_{max}$
of a shower
maximum (or the curve of the longitudinal development of a shower measured by the fluorescent method),
the values of signals in some various detectors
on the ground and signals in muon detectors. Again and in this case any conclusions are severe model dependent. So,
it is of primary importance to study also parameters of interactions of particles at ultra high energies. The energy
$E$ of the primary particles which induced a shower should also be known. There are some standard methods to estimate
the energy $E$ of a shower. But any alternative methods of energy estimation are also of interest. It was suggested
that readings of all detectors should be compared with calculated signals for a shower with the given values of the
zenith and azimuth angles \cite{13}. Calculations have been carried out for four showers observed at the Yakutsk array (YA)
\cite{14,15}. It should be mentioned that we use results of simulations for some sample of individual showers to
take into account fluctuations in the longitudinal and lateral development.
In this paper we use all readings of scintillation detectors placed on the ground, the muon underground detectors
and the detectors of the Vavilov-Cherenkov radiation. First, the energy of a shower and coordinates of its axis were
estimated with the help of the total signals in ground scintillation detectors. Then we repeated this procedure for
readings of detectors of the Vavilov-Cherenkov radiation to check the energy estimate found at the previous step. At
last, the muon detector readings have been used. At each step we try to draw some conclusions about chemical composition
of the primary cosmic radiation at ultra high energies.

\section{Method of simulations}

Simulations of the individual shower development in the atmosphere have been carried out with the help of the code
CORSIKA 6.616 \cite{16} in terms of the models QGSJET2 \cite{17} and Gheisha 2002 \cite{18} with the weight parameter
$\epsilon={10}^{-7}$ (thinning). The program GEANT4 \cite{19}
has been used to estimate signals in the scintillation detectors from the shower electrons, positrons, gammas
and muons at different points from the shower axis. For the same shower also at different points from the shower axis
signals in detectors of the Vavilov-Cherenkov radiation and in the muon detectors have been calculated. Simulations
have been carried out for four or two species of the primary particles (protons and nuclei of helium, oxygen and iron)
with a statistics of four individual events for every species of primaries. The energy $E$ of every shower was assumed
in calculations to be equal to value $E_{exp}$ estimated previously. Some sample of individual simulated showers induced
by various primaries particles have been constructed. These individual showers allow to take into account fluctuations
in the longitudinal development of a shower. Then the ${\chi}^{2}$ method has been used to find out which of calculated
individual showers agree best with data.

It was assumed in accordance with experimental data that the most energetic shower observed at the YA consists mainly of
muons and their deflections in the geomagnetic field have been taken into account. Readings of all scintillation
detectors have been used to search for the minimum of the function ${\chi}^{2}$ in the square with the width of 400~m and a
center determined by data with a step of 1~m.
These readings have been compared with calculated responces which were multiplied by the coefficient $C$. This
coefficient changed from 0.1 up to 4.5 with a step of 0.1.
Thus, it was assumed, that the energy of a shower and signals in the scintillation detectors are proportional
to each other in some small interval.
New estimates of energy

\begin{equation}
E=C\cdot E_{exp},~eV             \label{eq:2}
\end{equation}

\noindent
where coefficient $C$ shows the difference with the experimental estimate of energy $E_{exp}$, coordinates of axis and
values of the function ${\chi}^{2}$  have been obtained for each individual shower separately for total signals in scintillation
detectors, signals in detectors of the Vavilov-Cherenkov radiation and signals in muon detectors. The four extensive air
showers observed at the YA have been interpreted with the help of this calculations.

\section{Results of the study of the chemical composition}

 \begin{figure}[!t]
  \centering
  \includegraphics[width=8cm]{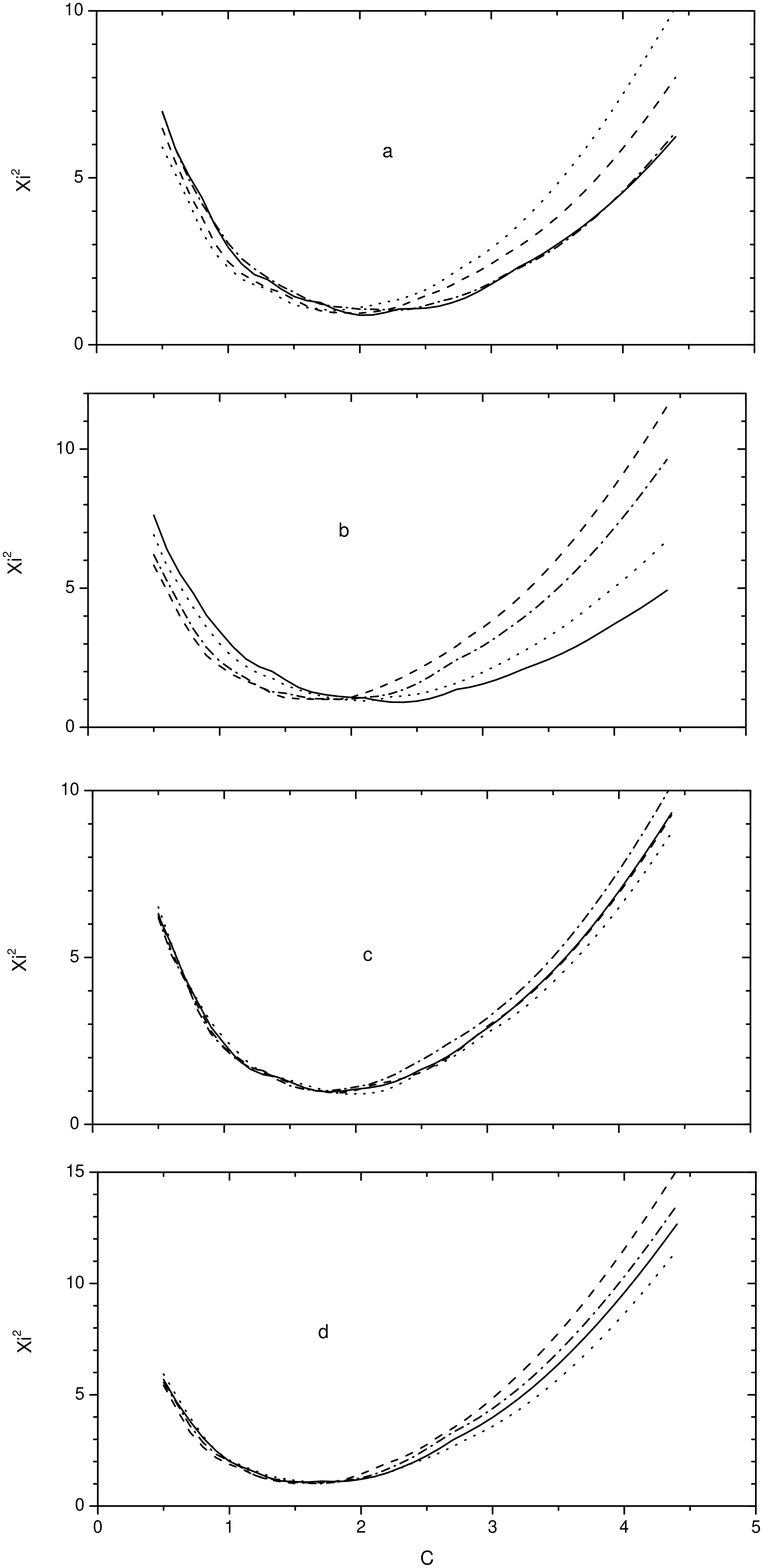}
  \caption{The values of the ${\chi}^{2}$ function per one degree of freedom vs the energy coefficient $C$ for various
  primaries: a -- protons, b -- helium nuclei, c -- oxygen nuclei, d -- iron nuclei}
  \label{Fig. 1}
 \end{figure}

\begin{figure}[!t]
  \centering
  \includegraphics[width=8cm]{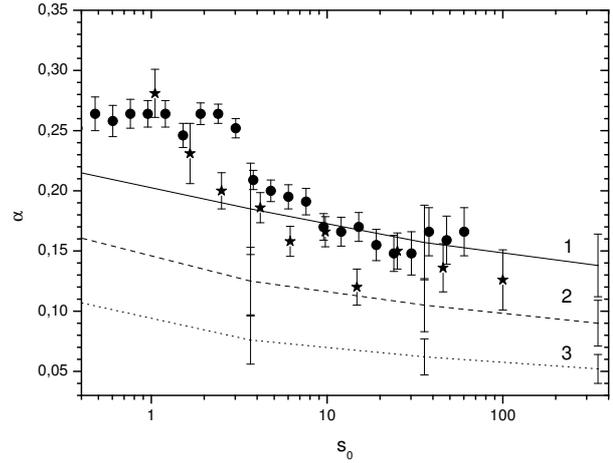}
  \caption{Fraction $\alpha$ of muon contribution to signal in a scintillation detector for the vertical extensive
  air showers. Points -- \cite{20}, stars -- \cite{21}. Calculated curves: 1 -- $E_{\mu}=0.3$~GeV,
  2 -- $E_{\mu}=1$~GeV, 3 -- $E_{\mu}=2$~GeV}
  \label{Fig. 2}
 \end{figure}

First, data of the most energetic shower have been interpreted. The 16 various values of energy estimates for 16
individual simulated showers with different
values of the function ${\chi}^{2}$ have been obtained for the same sample of the 31 experimental readings of
the scintillation detectors. Fig. 1 illustrates the dependence of the values of ${\chi}^{2}$ function per one
degree of freedom on the energy coefficient $C$ for four species of the primary particles and four events for every species
as follows: (a) -- for the primary protons, (b) -- for helium nuclei, (c) -- for oxygen nuclei and (d) -- for iron nuclei.
The systematic decreasing of the coefficient $C$ can be seen with increasing of atomic number of the primary particles
(from $\sim 2$ for the primary protons to $\sim 1.7$ for the primary iron nuclei). As the giant shower is very inclined muons give
main contribution to signals in the scintillation detectors. For the iron primaries which produce more muons than the
proton one the energy estimates are by a factor 1.3 -- 1.5 less than for the proton primaries. The values of the ${\chi}^{2}$
function per one degree of freedom are increasing from $\sim 0.9$ for the primary protons to $\sim 1.1$ for the primary iron nuclei.
Thus, all species of primary particles are possible for this particular shower. It should be mentioned that muons
contribute $\sim 80\%$ of the total signal in this inclined shower. Therefore, in this case readings of muon detectors do
not provide additional information relatively to the readings of scintillation detectors on the ground. So, it is of
important to find out contribution of muons to total signals for the vertical showers.
Fig. 2 shows the fractions of
the total signal which are contributed by muons at a distance of 600~m from the shower axis. The curves 1, 2 and 3
are calculated for muons with the threshold energies 0.3, 1 and 2~GeV accordingly. The experimental data \cite{20,21}
which are obtained for various zenith angles are also shown. These data were corrected to the vertical showers but
this correction is somehow uncertain. The data show that muon contribution to total signal decreases from $20\%$ at the
energy $E={10}^{18}$~eV to nearly $15\%$ at the energy $E=4\cdot {10}^{19}$~eV. The curve 2 calculated for the primary protons shows a
change from $\sim 13\%$ to nearly $\sim 9\%$ at the same energy interval. The difference is very large. It exceeds a factor of 1.6.
It is of primary importance. If data show correct values then the models QGSJET2 \cite{17} and Geisha 2002 \cite{18}
are unable to reproduce the
correct values of muon contribution to total signal for the primary protons. Only the primary iron nuclei may fit
the data. So, the conclusion about the energy estimates of the giant shower observed at YA \cite{14} should also be made
for the primary iron nuclei. Thus, chemical composition at ultra high energies can be studied in terms of these models.
To make it more definite we estimated energies of three more showers with the help of three methods. First, total
signals in the scintillation detectors have been used. Secondly, readings of detectors of the Vavilov-Cherenkov
radiation have been exploited. At last, the muon detector readings were compared with the appropriate calculated
signals. For the first shower with the energy $E_{exp}=6.5\cdot {10}^{19}$~eV coefficient $C$ obtained with the help of the first
method happened to be $\sim$ 0.5 -- 0.75 for the primary protons and $\sim 0.9$ for the primary iron nuclei with values of the
${\chi}^{2}$ function per one degree of freedom $\sim 2.5$ and $\sim 1.4$ accordingly. So, the primary iron nuclei have some privilege.
With the help of detectors of the Vavilov-Cherenkov radiation the appropriate coefficients $C$ were found to be $\sim 1.5$
for the primary protons and $\sim 1.1$ for the primary iron nuclei also with some privilege to iron nuclei. As for signals
in muon detectors these coefficients $C$ equal to $\sim 2.2$ and $\sim 1.6$ accordingly with the values of the ${\chi}^{2}$ function per
one degree of freedom  $\sim 0.5$ and  $\sim 0.8$. The similar results have been obtained for the second shower with the energy
$E_{exp}=2.5\cdot {10}^{19}$~eV. For the primary protons and iron nuclei the appropriate coefficients $C$ equal to $\sim 0.65$ and $\sim 0.9$
with the values of the ${\chi}^{2}$ function per one degree of freedom $\sim 2.5$ and $\sim 1.7$ for the first method. Second method
gave coefficients $C$ $\sim 1.1$ and $\sim 0.9$ for protons and iron nuclei accordingly. Coefficients $C$ $\sim 2.6$ and $\sim 1.6$ were
obtained with the help of muon detectors with the values of the ${\chi}^{2}$ function per one degree of freedom  $\sim 1.3$ and  $\sim 0.9$
accordingly. The third shower with the energy $E_{exp}=5\cdot {10}^{19}$~eV has the coefficients $C$ $\sim 0.75$ and $\sim 0.6$ with the values
of the ${\chi}^{2}$ function per one degree of freedom ~3.5 and ~3.0 for the first method. The second method gave coefficients
$C$ $\sim 1.1$ and $\sim 0.9$. Unfortunately, for this shower there no readings of muon detectors. New coordinates of shower axis
vary from the experimental one by some dozen of meters. So we may conclude that in terms of the QGSJET2 \cite{17} and Gheisha 2002
\cite{18} models heavy primaries such as iron nuclei have some privilege.

\section{Conclusion}

The three methods have been used to estimate energy of four extensive air showers by comparison different detector readings
with calculated signals in individual events for various primary particles with the help of the code CORSIKA 6.616 \cite{16} in terms of the
model QGSJET2 \cite{17} and Gheisha 2002
\cite{18} models with the weight parameter $\epsilon={10}^{-7}$ (thinning). The program GEANT4 \cite{18} has been used to estimate signals in the
scintillation detectors from the shower electrons, positrons, gammas and muons at different points from the shower axis. It was
found out that in terms of the QGSJET2 \cite{17} and Gheisha 2002
\cite{18} models heavy primaries such as the iron nuclei fit data for four showers better than
the primary protons. It was also stressed that any conclusions are very model dependent. Thus, to be more confident in results
of the study of chemical composition parameters of interactions of particles at ultra high energies and energy estimates of showers
should be known precisely.

\newpage

\section{Acknowledgements}

Moscow authors thank RFBR (grant 07-02-01212) and G.T. Zatsepin LSS
 (grant 959.2008.2) for support.\\
Autors from Yakutsk thank RFBR (grant 08-02-00348) for support.

\end{document}